\documentclass[conference]{IEEEtran}
\ifCLASSINFOpdf
\usepackage[pdftex]{graphicx}
\graphicspath{{./}}
\else
\fi

\usepackage{amsmath}
\usepackage{array}
\begin{document}
%
\title{Optimizing Performance on Trinity Utilizing Machine Learning, Proxy Applications and Scheduling 
Priorities
}

\author{\IEEEauthorblockN{Phil Romero}
\IEEEauthorblockA{High Performance Computing Division\\Los Alamos National Laboratory\\
Email: prr@lanl.gov}}



%


\maketitle

\begin{abstract}
Abstract—The sheer number of  nodes continues to increase in today’s supercomputers, the first half 
of Trinity alone contains more than 9400 compute  nodes. Since the speed of  today’s clusters are 
limited by the slowest  nodes, it more important than ever to identify slow nodes, improve their 
performance if it can be done, and assure minimal usage of slower nodes during performance critical 
runs. This is an ongoing maintenance task that occurs on a regular basis and, therefore, it is 
important to minimize the impact upon its users by assessing and addressing slow performing nodes 
and mitigating their consequences while minimizing down time.  These issues can be solved, in large 
part, through a systematic application of fast running hardware assessment tests, the application 
of  Machine Learning, and making use of performance data to increase efficiency of large clusters. 
Proxy applications utilizing both MPI and OpenMP were developed to produce data as a substitute 
for long runtime applications to evaluate node performance. Machine learning is applied to identify 
underperforming nodes, and policies are being discussed to both minimize the impact of 
underperforming nodes and increase the efficiency of  the system. In this paper, I will describe 
the process used to produce quickly performing proxy tests, consider various methods to isolate the 
outliers, and produce ordered lists for use in scheduling to accomplish this task.


\end{abstract}


%
\IEEEpeerreviewmaketitle

\section{INTRODUCTION}





Today’s supercomputers consist of ever larger and growing numbers of nodes, since computational 
tasks running on these computers are limited to progressing at  rates that are limited by the 
slowest performing components, it is more important than ever to identify bottlenecks and mitigate 
their damaging effects in a time efficient manner. There are several factors that contribute to 
producing the slow progression of computational tasks including:
\begin{enumerate}
\item
Slow performing CPU’s
\item
Slow performing Random Access Memory
\item
Slow performing and/or less than optimally buffered input/output systems
\item
Slow performing node interconnects.
\end{enumerate}

\begin{figure}
    \centering
    \includegraphics[scale=0.17]{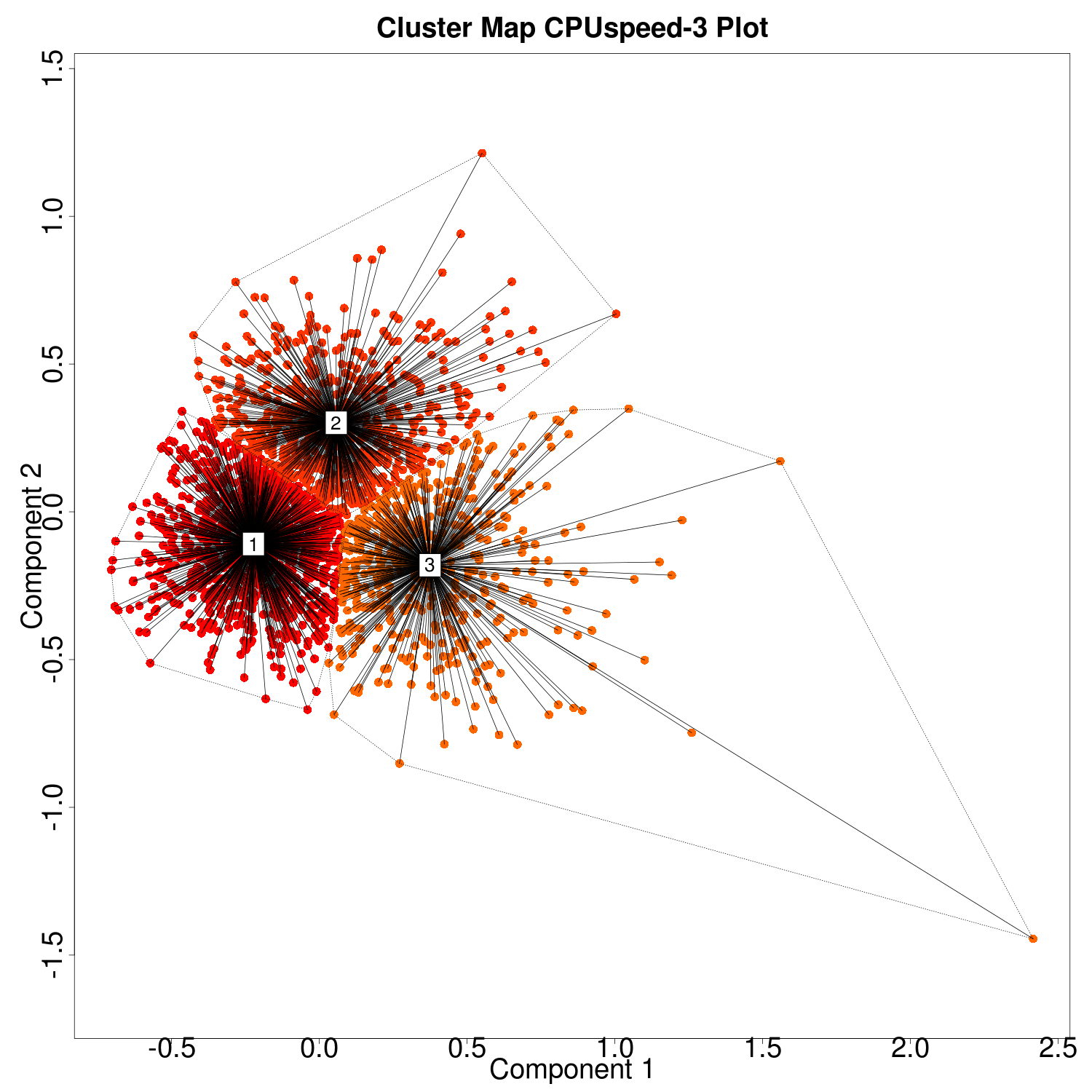}
    \caption{This figure shows a map plot that attempts to preserve distances between the features of all items under consideration.  Note the cluster labeled ”3”, it shows a markedly different shape 
from the clusters labeled ”1” with large outliers. The x and y axes represent distances in 
arbitrary
space and are orthogonal.}
    \label{fig:META_OPT_TRY11}
\end{figure} 

Factory tests allowed Los Alamos extensive time to test many different applications on a two 
thousand node subset of Trinity. Four different tests were conducted targeting each performance 
category of cpu speed, memory speed, and interconnect bandwidth. The results were then mined for 
clusters utilizing the kMeans clustering algorithm[1]. The output of the clustering process proved interesting in that cpu speed tests produced the largest variations in performance. This can be seen in Figure 1, the clusters labeled 2 and 3 show widespread outliers that produce irregular shape clusters as compared to the cluster labeled 1. This plot attempts to preserve distances in two dimensions that enable a better assessment of the variability in items than would be found through a Principal Components Analysis[4] that is sensitive to a given projection of data from many dimensions  into two dimensions.  Similar plots produced for memory speed and interconnect  bandwidth do not show widespread outliers, they produce relatively compact and regular shapes.This performance imbalance indicates that extra effort should be expended in reducing the differences in performance amongst cpu related tasks to increase efficiency in balancing the cluster, rather than expending efforts in minimizing interconnect and memory performance differences. 

Assessing underperforming nodes is often done by performance results produced by the High Performance Linpack[2] currently. This is done for a variety of reasons including that  is an accepted proxy for node performance that  approximates well the majority of applications that are typically run on our  clusters. However, with the advent  of larger memory capacities, the standard HPL, when tuned for near optimal performance, can produce run times in excess of four hours to properly assess a node. Also, random variability of performances needs to beconsidered when assessing the performance of nodes, making it necessary to have  more than just a few measurements in order to be confident that nodes are properly measured. Consequently, time considerations dictate that quickly running applications serve as a proxy for recognized performance standards such as the HPL. Once performance information for each node is found and validated, it can be utilized in various scheduling policies in order to increase the use efficiency of the cluster in a multitude of ways.

An outline of the proposed proxy applications is shown in Section II. Results obtained from the High Performance Linpack are given in Section III.  Mapping performance of the proxy applications to the HPL is discussed in Section IV. Mitigation strategies for underperforming nodes is discussed in Section V. Finally, a conclusion is presented in Section VI.

\section{METHODOLOGY, AN OUTLINE OF THE PROXY APPLICATIONS}


The High Performance Linpack is a convenient  software package utilized for assessing node 
performance for a number of reasons including: It produces performance numbers that are widely 
published and so comparables can usually be identified. It’s extensive optimization attempts to 
extract performance often pushing the limits of  the hardware. It utilizes a good portion of the 
memory per node, thereby also serving as a memory test and not only a cpu test. The results it 
produces are generally useful to commonly run applications  on our supercomputers, save for the 
very sparse matrix problems which would be covered by the HPCG[3].

However, in order to extract high levels of performance, approximately eighty percent of the memory 
must be utilized. This necessitates larger matrices to be solved as the memory capacity of a node 
increases.  The larger order matrices necessarily produce longer run times for completion 
necessitating more than 26 minutes in run time for   a single HPL result since Trinity nodes are 
equipped with 128GB. A good solution would be to find applications, or write them, that will mimic 
the performance numbers that would be obtained by the HPL (a performance distribution is shown in 
Figure 2) and provide many performance measurements per run, thereby allowing for the assessment of 
performance variability, in a small fraction of the run times needed by the HPL.

Modern nodes utilized in supercomputers contain many cores, each Trinity node is composed of Intel Xeon Phi 7250 (KNL) processors, each having 68 cores. In order to properly assess performance for these nodes it is imperative that all cores be tested simultaneously. There are two dominant methods at this time for writing applications that utilize parallelism well, the first is MPI and the second is OpenMP. It is proposed that short algorithms measured very quickly and repetitively could provide the results to properly assess the performance of nodes that identify the same outliers that the HPL test would identify. Relaxed HPLs have had little correlation between optimized performance and small order matrix input results. Therefore, it is hypothesized that tests that push performance to extreme limits maybe a better indicator of node performance than detuning applications to run in shorter periods of time.The algorithms implemented for this purpose are outlined below.

\begin{figure}
    \centering
    \includegraphics[scale=0.20]{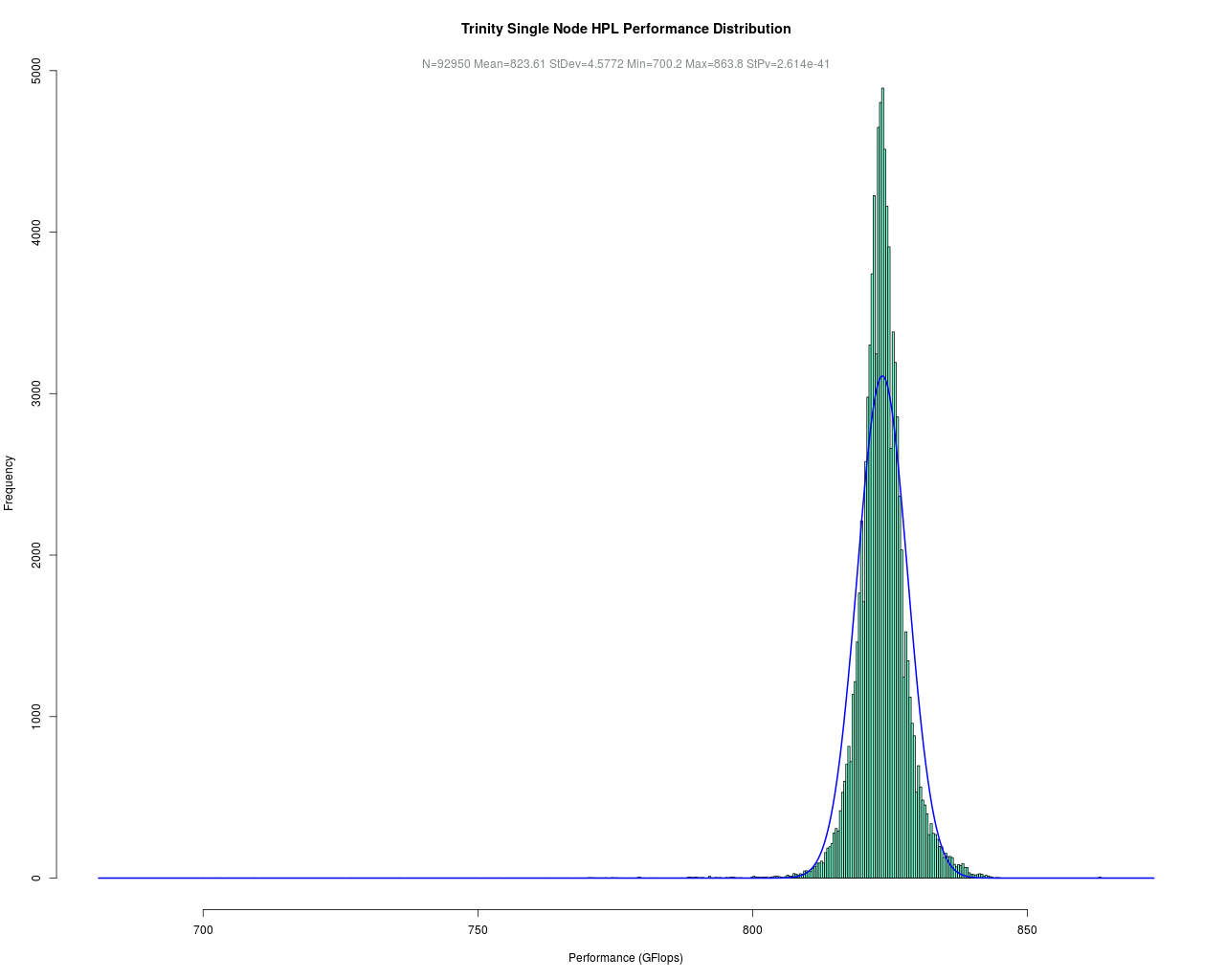}
    \caption{This figure shows a performance distribution of the HPL,  note the distribuition is 
similar to the blue line indicating a normal distribution.} 
    \label{fig:RAWSIGSAMPLE1}
\end{figure}

\begin{figure}
    \centering
    \includegraphics[scale=0.20]{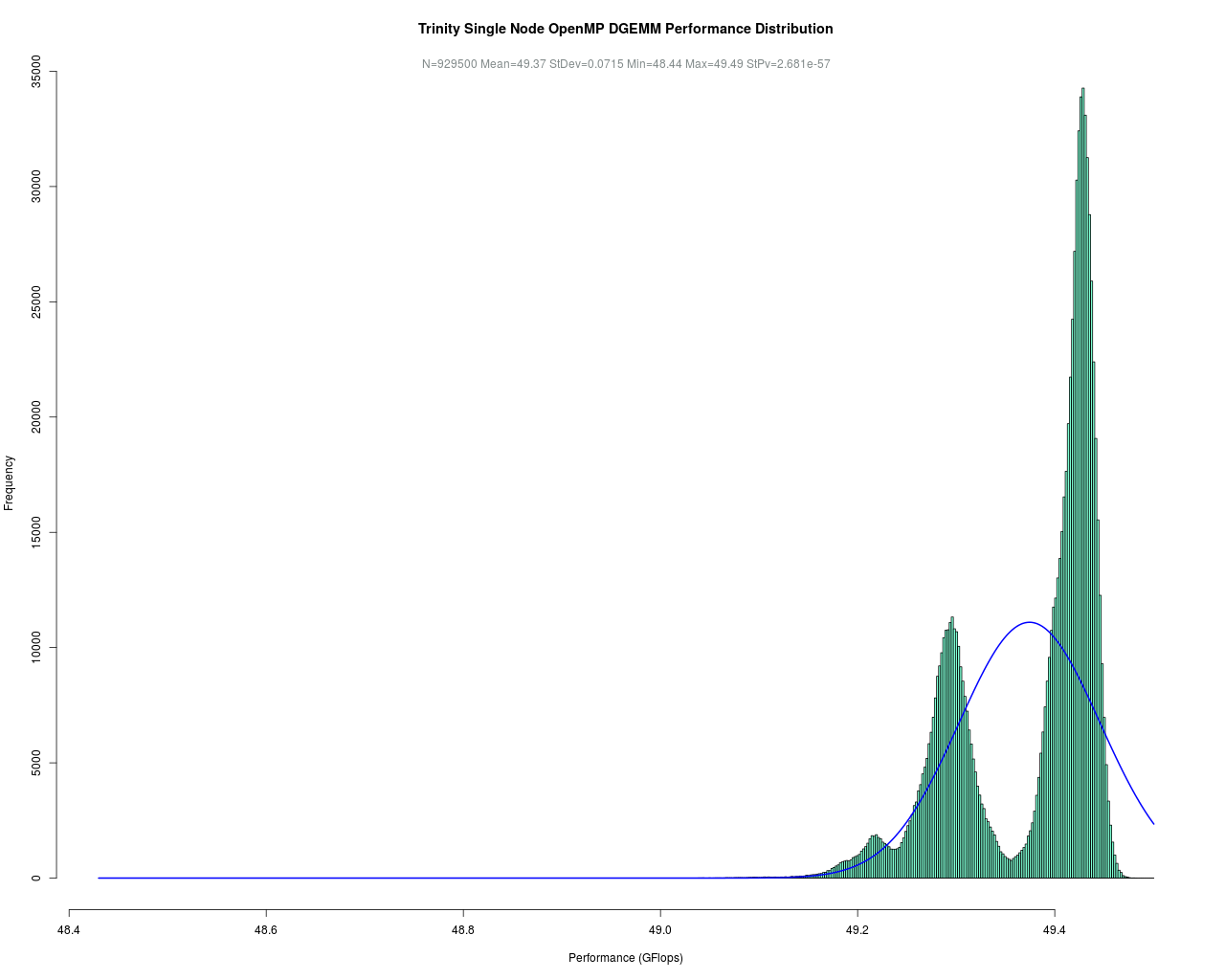}

    \caption{This figure shows a performance distribution of the OpenMP DGEMM implementation. Note 
that this is not a singular peak distribution, three peaks are discernible, consequently it may be 
expected to segregate node performances better than an algorithm that produces a singular peak
distribution.}
    \label{fig:RAWSIGSAMPLE2}
\end{figure}

All applications outlined below are either an OpenMP or MPI based application, each provides performance measurements not only by node, but also by core, allowing for the assessment of what core, or cores, is/are responsible for the performance degradation.

The first application is an OpenMP based DGEMM implementation that utilizes the fused multiply add instructions newly available on the Intel KNL processors. A performance distribution for this algorithm is shown in Figure 3.

The second application is an OpenMP based DGEMM implementation that utilizes the intel mkl library cblas dgemm function call. This application produces a near normal distribution is not shown.

\begin{figure}
    \centering
    \includegraphics[scale=0.20]{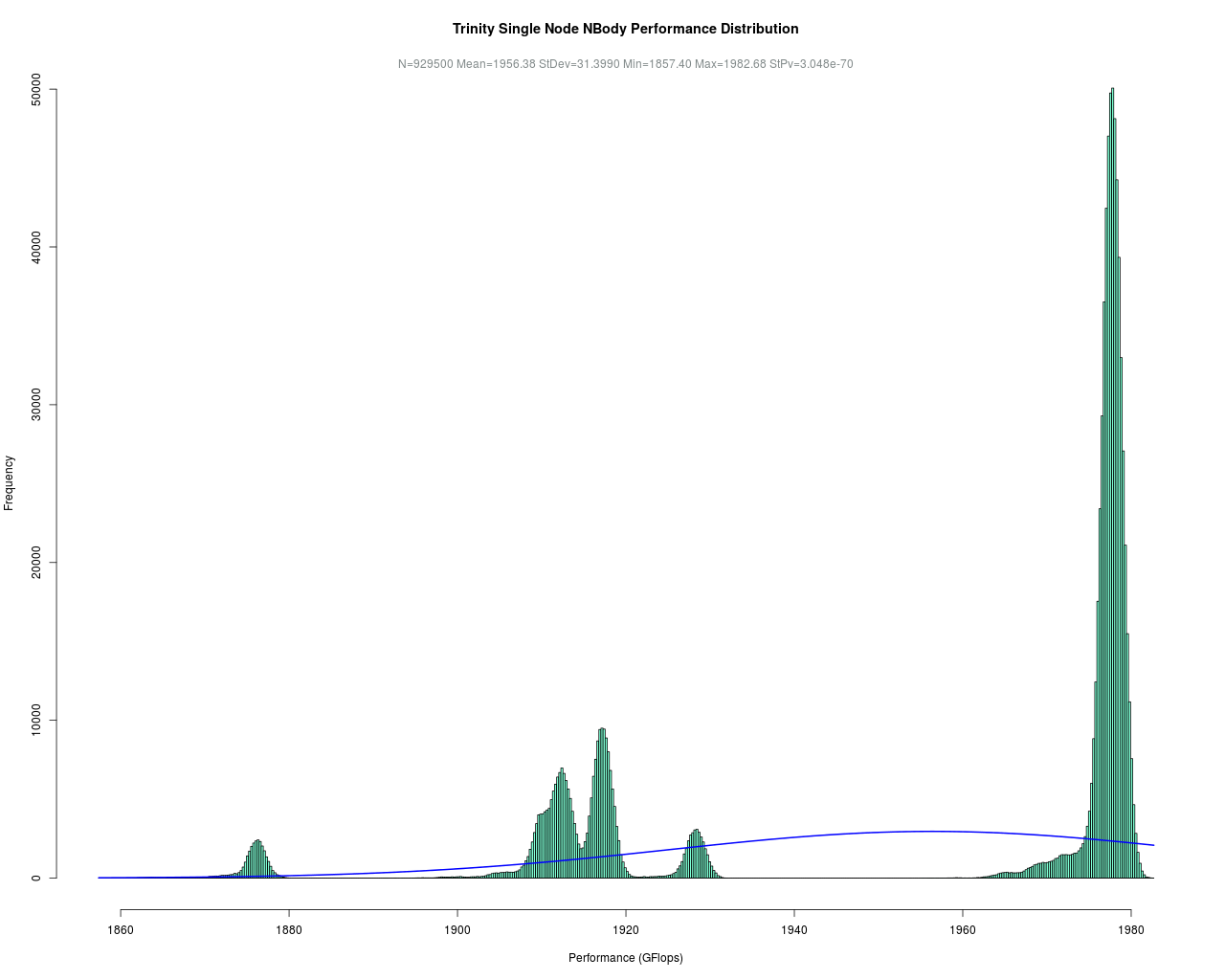}

    \caption{This figure shows the performance distribution of the OpenMPbased NBODY algorthithm that utilizes fused multiply add instructions newly available for the KNL architecture. Note that the distribution has five peaks instead of just one.}
    \label{fig:RAWSIGSAMPLE3}
\end{figure} 

\begin{figure}
    \centering
    \includegraphics[scale=0.20]{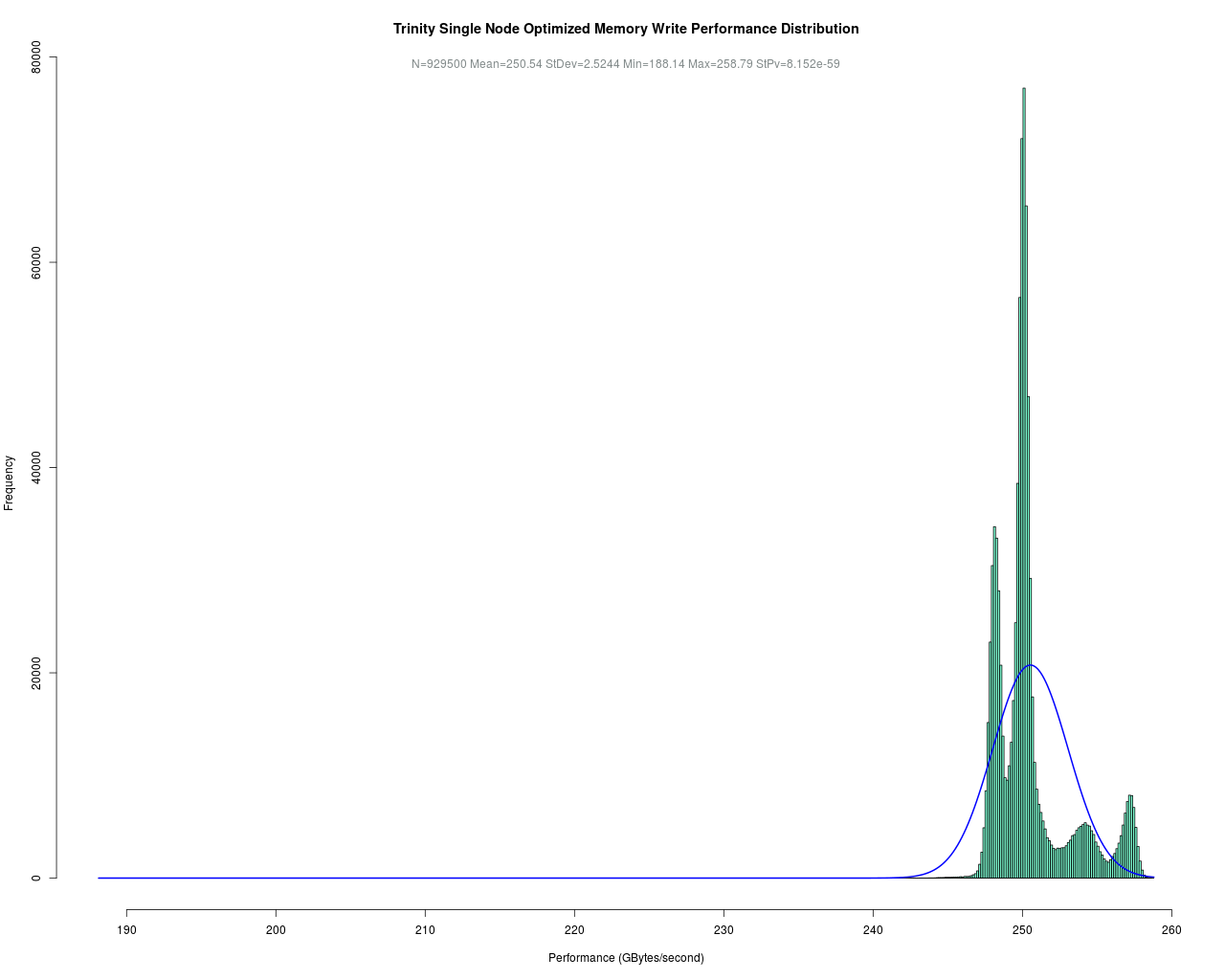}

    \caption{This figure shows the performance distribution of an optimized memory write performance test, note there arefour peaks in the distributionand a long tail exists.}
    \label{fig:RAWSIGSAMPLE4}
\end{figure}

The third application is an OpenMP based NBODY implementation that utilizes the fused multiply add instructions newly available on the Intel KNL processors. A performance distribution for this algorithm is shown in Figure 4.

The fourth application is an OpenMP based unoptimized memory test implementation, it times writes from each core that target 120 GB of the 128 GB on each node.  This application produces a near normal distribution with a fairly long tail on the lower performance side and is not shown.

The fifth application is an OpenMP based optimized memory test implementation, it times writes from each core that target 120 GB of the 128 GB on each node.  A performance distribution for this algorithm is shown in Figure 5.

The sixth application is an MPI based implementation of the DGEMM algorithm. This application produces a near normal distribution with a slight leftward lean and is not shown.

The seventh application is an MPI based implementation of the DGEMM algorithm. This application also produces a near normal distribution and is not shown.

\begin{figure}
    \centering
    \includegraphics[scale=0.20]{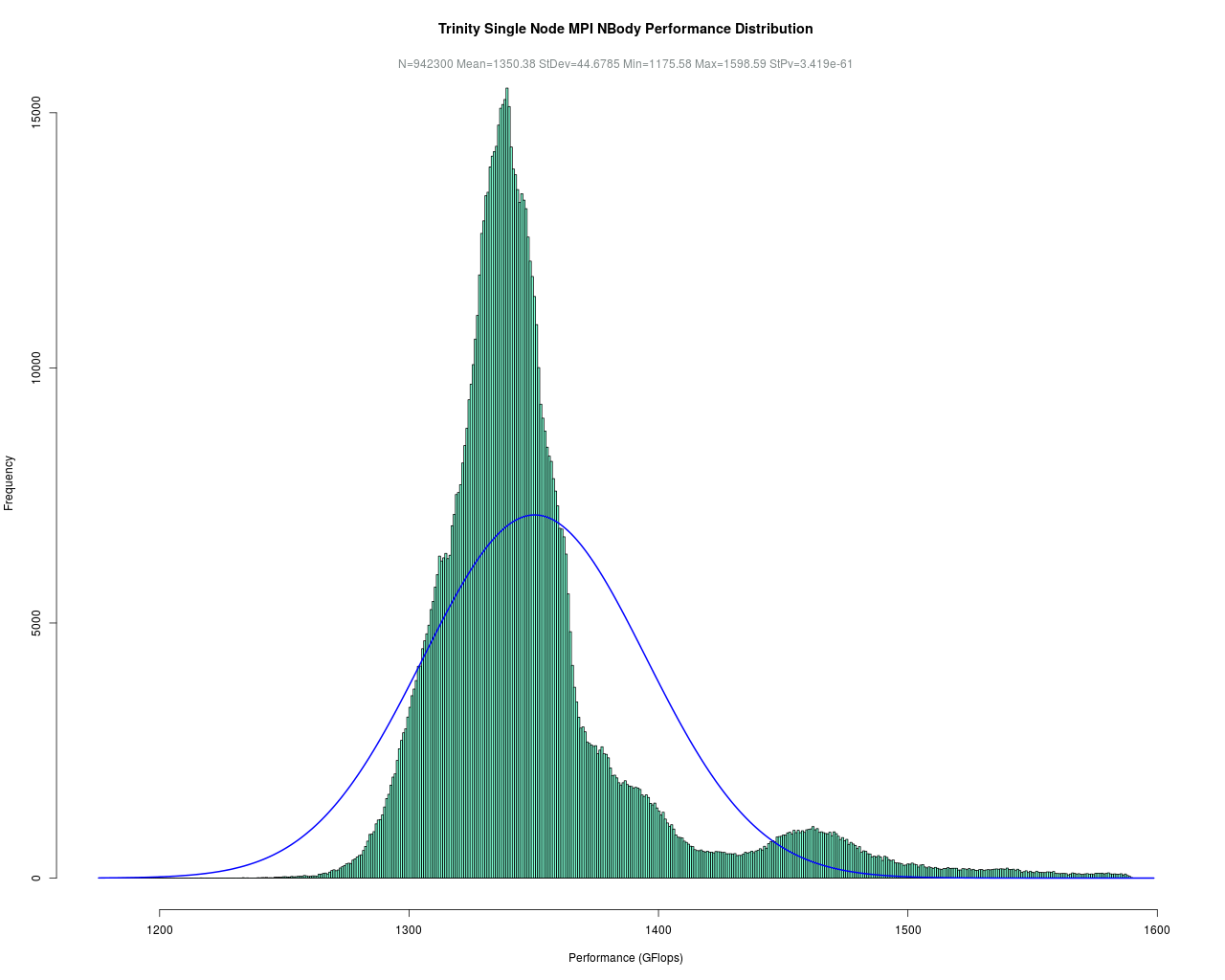}
    \caption{This figure shows the performance distribution of the NBODY
algorithm that utilizes the fused multiply add instruction available on the
KNL architecture. Note that the distribution has two distinct peaks.}
    \label{fig:RAWSIGSAMPLE5}
\end{figure} 

The eighth application is an MPI based implementation of the NBODY algorithm. It utilizes the fused multiply add instruction available on the KNL architecture. A performance distribution for this algorithm is shown in Figure 6.  

The ninth application is  an MPI based implementation of an LU Factorization algorithm.  The HPL algorithm is composed of an LU Factorization and a subsequent DGEMM. The thought here is to determine if an algorithm decomposition provides better correlations for the HPL algorithm.  This algorithm produces a bimodal distribution with similar peaks and is not shown here.

The tenth application is  an MPI based implementation of a DGEMM that is also unoptimized, similar to the LU Factorization algorithm. It also produces a bimodal distribution with similar peaks and is not shown here.

Finally, a modified version of the HPL (Deflated HPL) was created to run very quickly, although the results do not correlate well with the fully optimized HPL, it is another data set that can be utilized to identify outliers.



\section{RESULTS OF HPL PERFORMANCE TESTS}


Performance results were processed in sets of  10 results from 10 to 50 results per node.  There were a total of 9327 nodes for which results were obtained. I am primarily interested in identifying the lowest performing nodes, the boxplots for 10 and 50 results are shown below in Figure 7 and Figure 8 respectively.  Notice the low performing nodes increase variance noticeably with increasing numbers of samples per node,  whereas there is no significant increase discernible amongst the middle eleven or top eleven nodes.  This prompts speculation that maybe the lowest performing nodes are also the nodes with the highest variance. Should this relationship hold, performance is likely to fall off more than expected when slow nodes are in use with high node count computational tasks.

\section{MAPPING PERFORMANCE RESULTS OF THE PROXY APPLICATIONS TO HPL PERFORMANCE}

Data for all proxy applications on every node was collected, statistically analyzed to find correlations between the performance numbers. R was used to perform a linear regression between all performance results of the proxy applications and the optimized HPL results. This was then reduced to a minimal set of tests necessary to assure the identification of slow nodes. This function was found to be:

\begin{center}
\textit{
((4MPI DGEMM Min)+ \\
 (2MPI DGEMM Mean)+ \\
  MPI NBODY Mean) \textless{} 7190.0}
\end{center}

Note that performance outliers occur below an x value of 7190 and above 7600 x values. The outliers below x values of 7190 are more than 3.5 standard deviations below the mean performance. This serves to identify 12 nodes thatare performing at least 3.5 standard deviations below the mean. It also identifies the lone high performance outlier, although no mitigation actions are necessary for the nodes that perform significantly above the mean. Finally, notice that all the remaining outliers occur within a range of performance where the vast majority of HPL performance means are well above the three standard deviations below the mean cutoff for low performing nodes. This also means that simple regressions cannot be done to separate outliers with this proxy application measurement. Also, none of the remaining proxy application results can be utilized to find a regression line that could be utilized to find any remaining outliers.  Performance outliers can also be  found by computing the Mahalanobis Distance utilizing multivariate data. This technique is best explained in two dimensions by plotting two performance variables for each item, nodes, in this case. The first performance data (feature) is plotted using the x axis and the second performance data is plotted against the y axis to produce a single point on a two dimensional graph. This is repeated for all of the nodes in the data set, typically this produces an dense cloud of an elliptical shape with data points that are isolated. The Mahalanobis distance is then calculated that measures how far a given data point is from the center of mass of the elliptical point cloud with the equation. 
  
\begin{figure}
    \centering
    \includegraphics[scale=0.35]{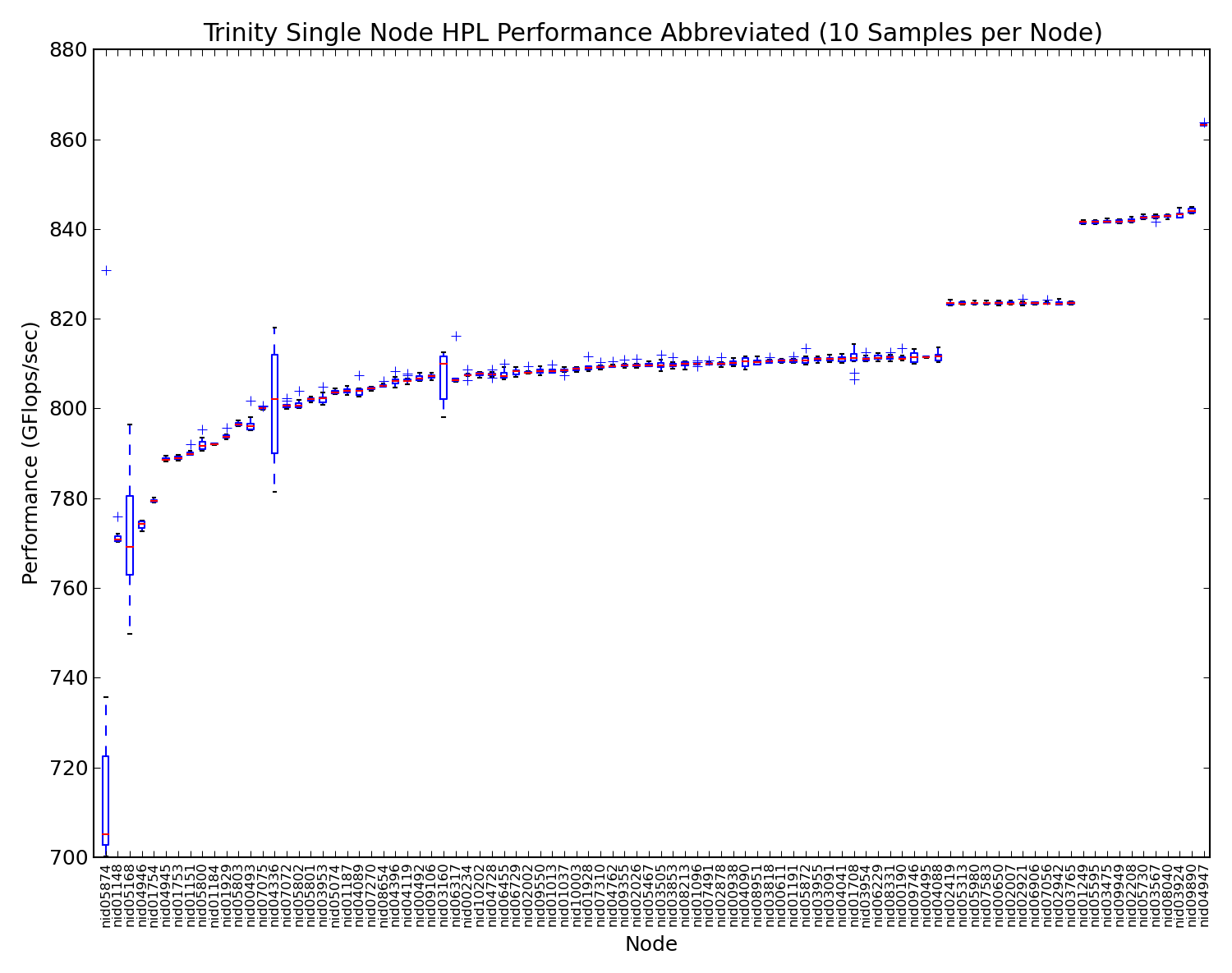}
    \caption{This figure shows a boxplot showing the lowest 70 performing
nodes (by mean performance), followed by the middle 11 performing nodes
and the top 11 performing nodes. This plot is for only 10 samples produced
from the HPL.}
    \label{fig:META_OPT_TRY1}
\end{figure}

\begin{figure}
    \centering
    \includegraphics[scale=0.35]{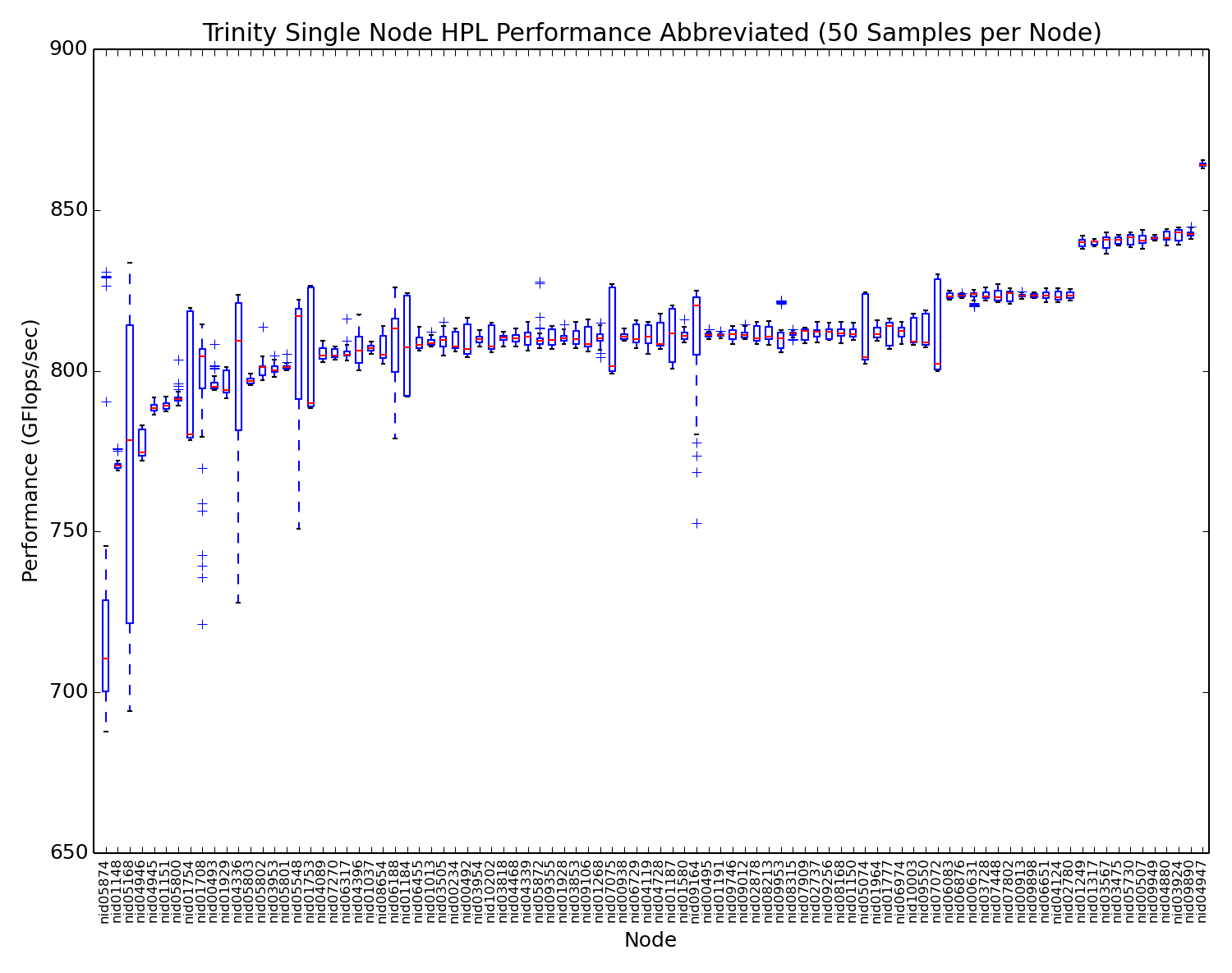}
    \caption{This figure shows a boxplot showing the lowest 70 performing
nodes (by mean performance), followed by the middle 11 performing nodes
and the top 11 performing nodes. This plot is for 50 samples produced from
the HPL.}
    \label{fig:META_OPT_TRY2}
\end{figure}

\begin{figure}
    \centering
    \includegraphics[scale=0.18]{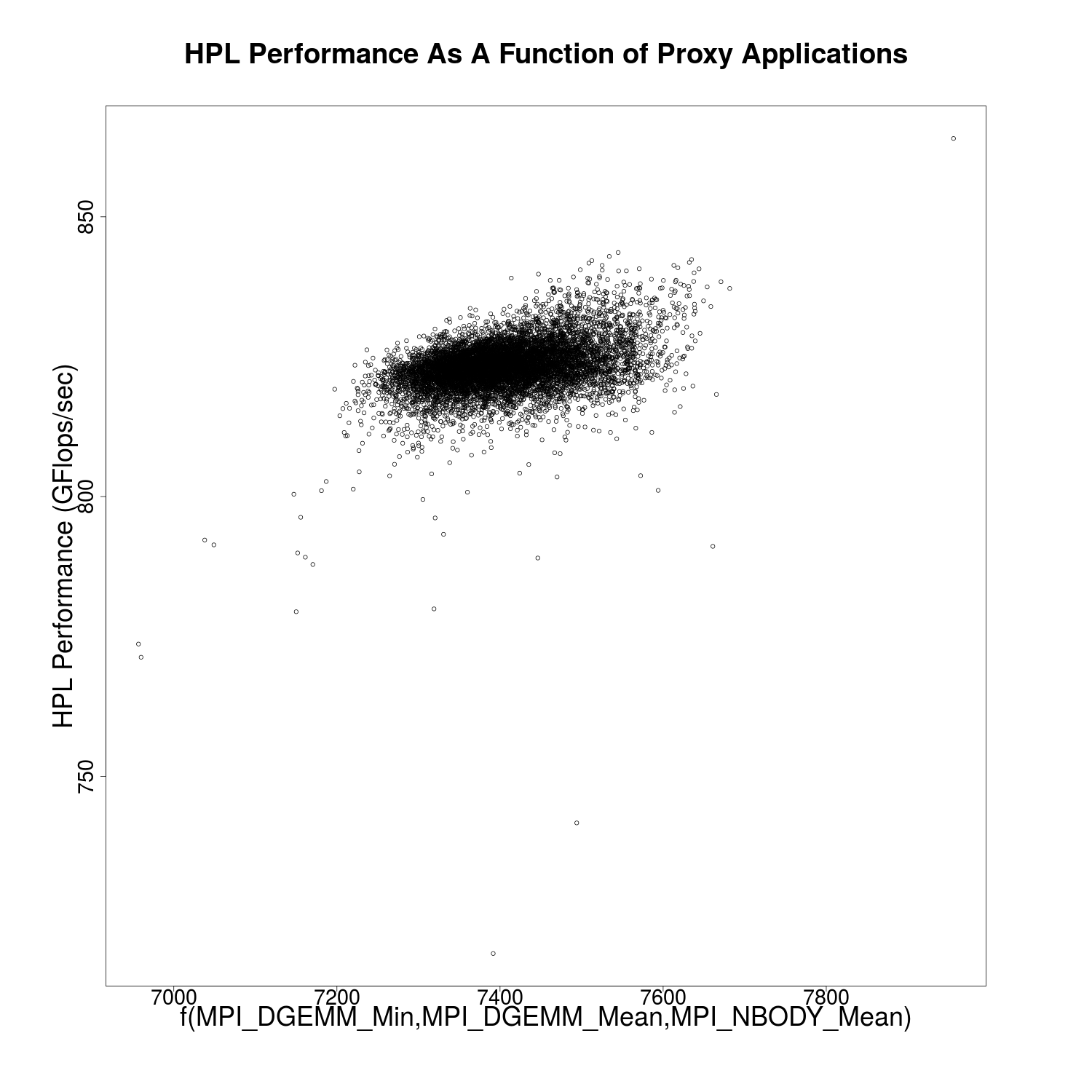}
    \caption{This figure shows a boxplot showing the lowest 70 performing nodes (by mean performance), 
followed by the middle 11 performing nodes and the top 11 performing nodes. This plot is for 50 
samples produced from the HPL.}
    \label{fig:META_OPT_TRY3}
\end{figure}


$$D_M = \sqrt{(X - M)^T S^{-1}(X - M)}$$

Where:
- $D_M$ is the Mahalanobis distance.
- $X$ represents a given data point.
- $M$ represents the mean of the dataset.
- $S$ represents the covariance matrix of the dataset.



The results for this function are shown in Figure 9. Note that performance outliers occur below an x value of 7190 and above  7600 x values.  The outliers below x values of 7190 are more than 3.5 standard deviations below the mean performance. This serves to identify 12 nodes that are performing at least 3.5 standard deviations  below the

It is important to note that Mahalanobis distance is not limited to two dimensions, it can be calculated in any number of dimensions. The Mahalanobis distances were calculated between all pairs of features (proxy application results) including HPL features, repeated this for all triplet pairs, and so on until all Mahalanobis distances were found up to all  sextuplet pairings. I then compared lists of  all the Mahalanbis distances greater than 3.5 standard deviations below the mean. The highest number of outliers that could be explained by this method was between the HPL Mean and the MPI DGEMM Mean values, the Mahalanobis Distance plot for these pairs is shown in Figure 10. The confusion matrix for this test of outliers is shown in Table I.

\begin{table}[htbp]
\centering
\caption{Confusion Matrix for Mahalanobis Outliers Between HPL Mean and MPI DGEMM Mean}
\label{table:my_table1}
\begin{tabular}{|c|c|c|}
\hline
  & Positive Label & Negative Label \\ \hline
Predicted Positive & 33 & 3 \\ \hline
Predicted Negative & 2 & 9245 \\ \hline
\end{tabular}
\end{table}

\begin{table}[htbp]
\centering
\caption{Mahalanobis Outliers Between Proxy Application Pairs}
\label{table:my_table2}
\begin{tabular}{|c|c|c|c|c|c|}
\hline
TP & FP & FN & TN & ProxyApp1 & ProxyApp2 \\ \hline
18 & 8 &  15 & 9279 & MPI LUFac Min &   Deflated HPL Mean \\ \hline
17 & 7 &  16 & 9278 & MPI DGEMM Min &  Deflated HPL Mean \\ \hline
20 & 65 & 13 & 9281 & MPI LUFac Mean &  Deflated HPL Mean \\ \hline
16 & 3 &  17 & 9277 & MPI DGEMMd Min & Deflated HPL Min \\ \hline
14 & 9 &  19 & 9275 & OMP NBODY MIN &  MPI LUFac Min \\ \hline
18 & 82 & 15 & 9279 & MPI LUFac StdDev &  MPI DGEMM Min \\ \hline
\end{tabular}
\end{table}

\begin{table}[htbp]
\centering
\caption{Error Rates in Percentages for Various adabag Package Settings}
\label{table:my_table3}
\begin{tabular}{|c|c|c|}
\hline
  & With Boosting  & Without Boosting  \\ \hline
Data 1 & Data 2 & Data 3 \\ \hline
Brieman Learning & 0.56  &    0.50 \\ \hline
Freund Learning  & 0.01  &     0.49 \\ \hline 
\end{tabular}
\end{table}

\begin{figure}
    \centering
    \includegraphics[scale=0.19]{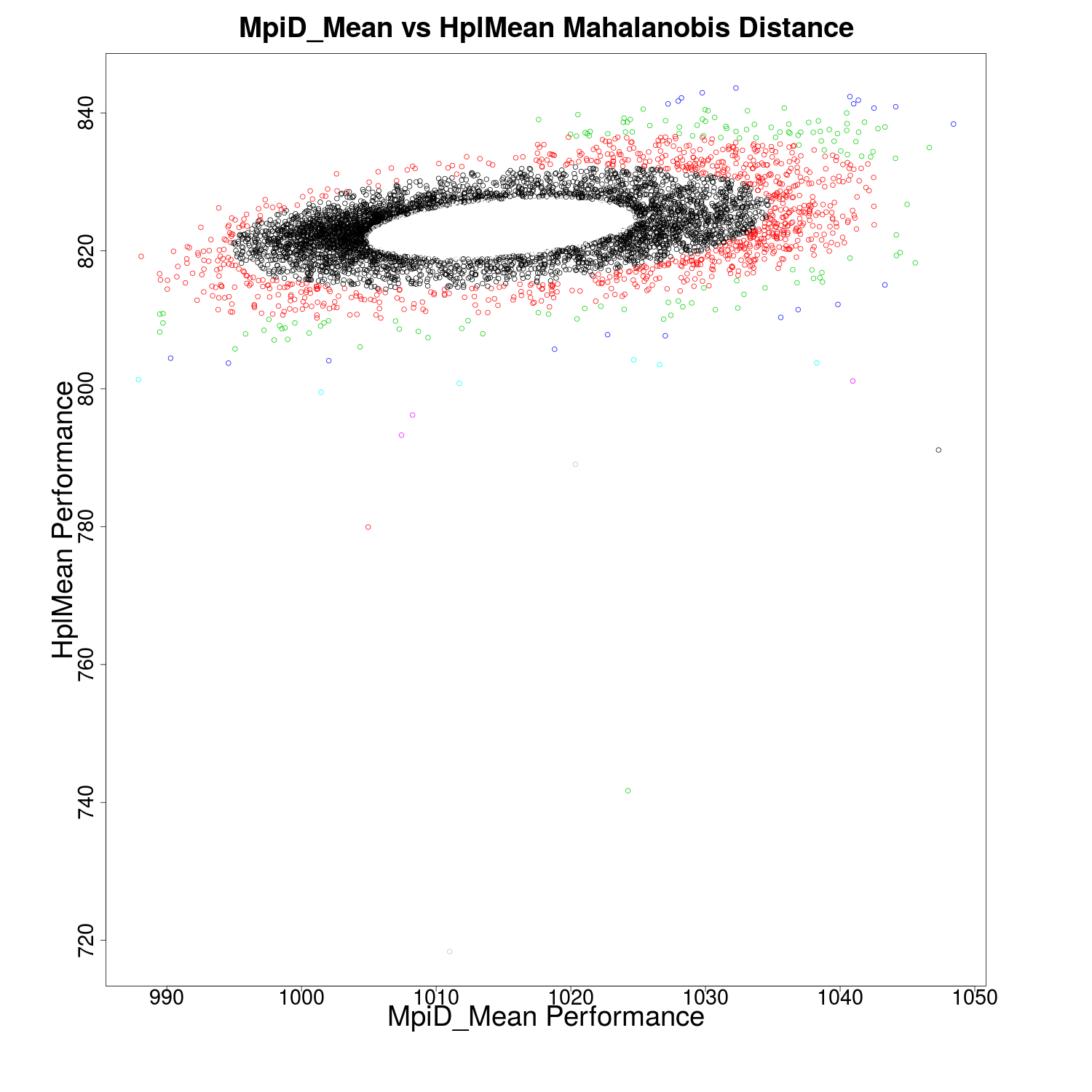}
    \caption{This figure shows a plot of MPI DGEMM Mean performance
versus HPL mean performances in GFlops/second, colors depict ranges of
Mahalanobis distances. black 1 to 2,red 2 to 3,green 3 to 4,blue 4 to 5,cyan
5 to 6 ... Note that outliers occur not only below about 804 GFlops/second
but also above 860 GFlops, these are outliers faster than the HPL mean
performance value that need not be attended to in terms of performance
mitigation.}
    \label{fig:META_OPT_TRY4}
\end{figure}

It should be noted that  although the error rate is low, this is not the best solution available because it assumes the HPL Mean is known. However, a large cluster should always be screened for slow nodes before it  is put into production, therefore this information should always be available. The negative side to this  is that it would be better  to have a complete replacement that runs in a much shorter  time that would approximate HPL Mean performance value. The proxy applications can be used to find slow nodes by finding the Mahalanobis Distance Outliers. Since more than 2k Pairs of Mahalanobis Outliers were calculated, there are many outlier pairs to select from, a very small sample of the outliers are shown in Table II. This table shows that at least 20 of the 33 outliers can be found with a single proxy application pair, since different pairs of applications often produce different outliers, it is entirely possible that much more than 20 of the 33 outliers will be found with this method.

There are many tools in Machine Learning that can be applied to this problem, the next tool considered is Random Forests. Random forests are a combination of tree predictors such that each tree depends on the values of a random vector sampled independently and with the same distribution for all trees in the forest[8]. This tool is used to classify labeled data, however, since there is a severe class imbalance,  it does not work well on this problem. The split on good nodes to slow 
nodes here is 9294 to 33 or about 0.35\%, very far from the 35\% to 65\% split where these algorithms work well. In recent years there have been many proposals as to how to treat class imbalance to allow for classification algorithms to provide better predictions[5]. Boosting the capability to make predictions is typically done by decreasing the number of negative samples in the data to produce a better balance, creating synthetic positive samples from the existing data samples and 
adding them to the data and/or using both to achieve a better balance between the classes. A popular package in R that addresses class imbalance is adabag[6], this package was used to see if an acceptable result could be produced. Low error rates were produced with this package and are shown in Table III. Although the Freund learning method with boosting provided the lowest error rate,it only predicted one true positive in an 83\% sample size of the tests while producing 40 false positives and 0 false negatives. This wasn’t significantly better than the Brieman learning method that predicted only 5 true positives in 83\% sample size of the tests while producing 28 false positives and 7 false negatives. The Mahalanobis distance outlier prediction produces less false predictions, both positive and negative, with more more than six fold the true positive rate thereby makingit a superior prediction system.

The boosting process can be used as input  into a deep neural network. The Mocha[9] package for Julia was selected to produce and run a deep neural  network of two hidden layers with 300 nodes in the first layer and 40 nodes in the second hidden layer. Many papers have been written as to how to deal with class imbalance including [10] but none of the recommendations worked in this case. It was decided to use the neural network for regression rather than classification.  This was done by boosting all  samples more than three standard deviations below the mean with both real observations and observations produced by the SMOTE method[11]  by over 300\%. The results of the nonlinear regression produced by the neural network is show in Figure 11. The horizontal andvertical lines serve to indicate the outliers in the lower  left quadrant. Although the results don’t look very useful  for predicting an exact value from the regression, it is extremely useful in assigning a probability that a node is an outlier, thereby augmenting the Mahalanobis Distance Outlier results to improve prediction accuracy. This process is ongoing at this time and is expected to produce results in a short  period of time.

\begin{figure}
    \centering
    \includegraphics[scale=0.18]{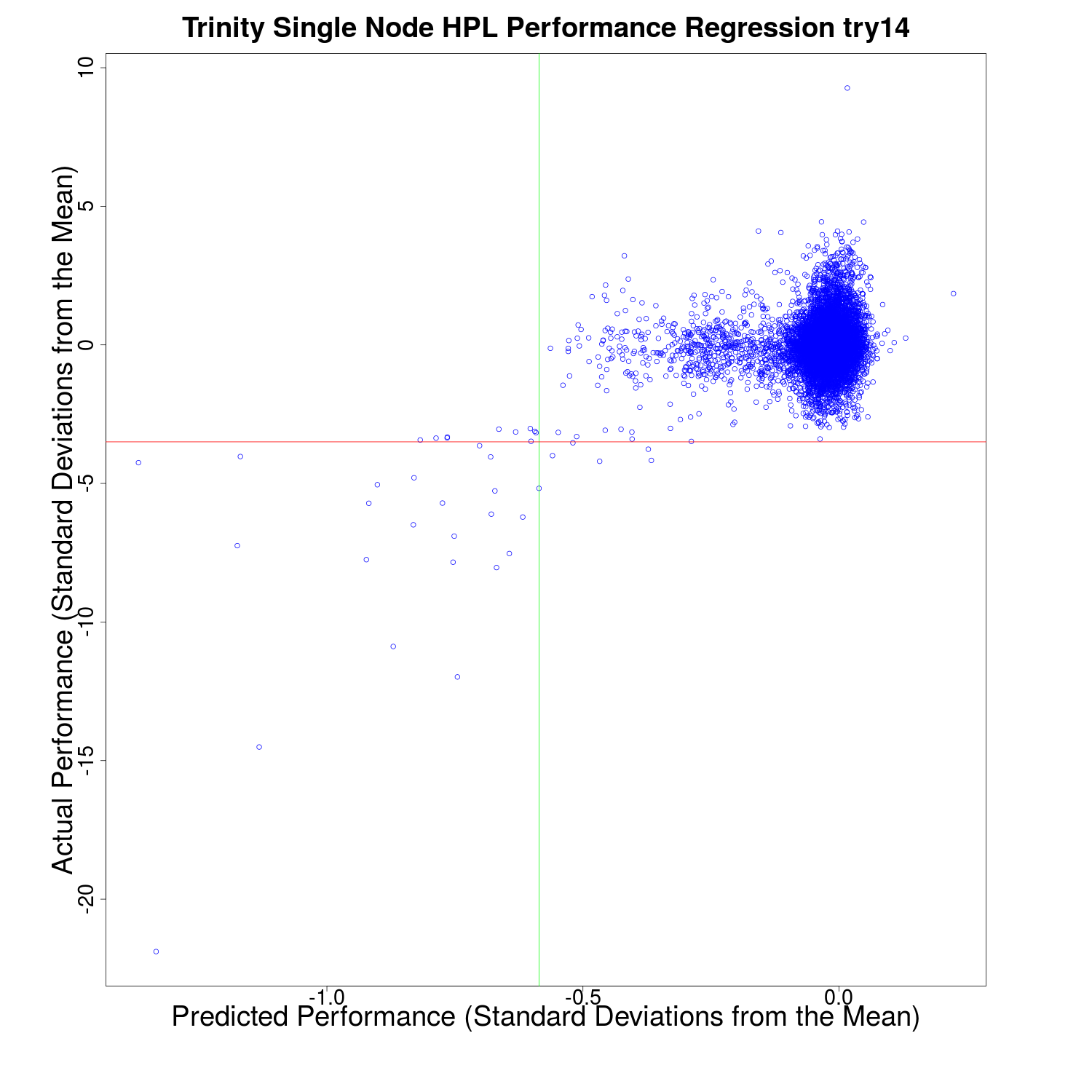}
    \caption{This figure shows the results of the nonlinear neural network
regression. All 9327 nodes are shown and the oultiers are demarcated in the
lower left quadrant (demarcated by the horizontal and vertical lines).}
    \label{fig:META_OPT_TRY6}
\end{figure}

\section{MITIGATION STRATEGIES FOR UNDERPERFORMING NODES}


Several strategies are being considered to mitigate  the performance consequences of slow nodes 
including:
\begin{enumerate}
\item
Applying a process called Trimming that can bring an underperforming node up to specification.
\item
Replacing the slow nodes outright.
\item
Assigning the slightly slow nodes to interactive tasks possibly including visualization, 
debugging, and compiling.
\item
Placing the slow nodes  in a performance ordered queue where they will be least likely to be 
utilized by large production jobs and most likely to be used by small batch jobs.
\end{enumerate}

There may also be possibilities of tuning the speed of nodes that an individual/project will be assigned by setting priorities for the highest performing nodes without adding any constraint that often becomes a performance bottleneck affecting more than just a few of the machines users. This could be done by altering the  ordering of the nodes by performance to favor certain projects/individuals/tasks over others in order to achieve maximum efficiency. It is also likely that nodes that may be slow for a fluid dynamics code may not correlate well with nodes that are slow for nbody calculations.  This would then open the door to highly customized node preference lists, again, towards maximizing efficiency.

\section{CONCLUSION}

A process has been described here that uses several different methods to identify slow nodes that would drastically reduce the performance of a cluster, in fact, I have identified a node that is at least 18\% slower on average than the mean performing nodes. This is equivalent to losing approximately 1678 nodes of a 9.3k node cluster. Although it is an ongoing process to be optimized, I was able to find twelve of the 33 slow nodes by simple regression from proxy applications,identify at least 20 of the 33 nodes by applying Mahalanobis distance techniques to identify the slow nodes. A neural network regression also clearly identifies 18 of the 33 outliers and can be used to augment the results of the other successful techniques to identify all of the bottom two thirds of true outliers, leaving only outliers that are only marginal. The huge class imbalance of having less than 0.35\% of nodes being slow is a major impediment to using Random Forest techniques, as is the feature space not being linearly separable. Further possibilities of how to improve the process have been identified and are being pursued.



\section*{Acknowledgments}

This manuscript has been approved for unlimited release and has been assigned LA-UR-23-29632.  This research used resources provided by the Los Alamos National Laboratory Computing Program. Los Alamos National Laboratory is operated by Triad National Security, LLC, for the National Nuclear Security Administration of U.S. Department of Energy (Contract No. 89233218CNA000001).  The author would like to thank Craig Idler who almost single handedly wrote the test harness named pavilion that allowed for collection of the more than five hundred million results produced for this study alone. Much appreciated is Dave Morton who made available dedicated time on the full cluster.  I also wish to thank Jennifer Green, Jim Williams and Nathan Hjelm who all shared cluster time.



%

\end{document}